\documentclass[a4]{article}
\usepackage{graphicx} 
\usepackage{cite}     
\usepackage{amsfonts,amstext,amssymb,amsmath}

\def \d {{\rm d}}
\def \e {e}
\def \TT {{\text{\texttt{T}}}\hspace{-0.04cm}{\text{\texttt{T}}}}
\newcommand{\Ibar}{\mathcal{I}\mkern-10mu{\text{{\Large -}}}{\,}}

\begin{document}

\title{The damping of gravitational waves in dust}

\author{Otakar Sv\'{\i}tek\\
Institute of Theoretical Physics, Charles University in Prague\\
Faculty of Mathematics and Physics\\
V~Hole\v{s}ovi\v{c}k\'ach 2, 180~00 Praha 8, Czech Republic}

\date{}
\maketitle

\noindent
{\bf e-mail address:} ota@matfyz.cz

\begin{abstract}
 We examine a simple model of interaction of gravitational waves with matter (primarily represented by dust). The aim is to investigate a possible damping effect on the intensity of gravitational wave when passing through media. This might be important for gravitational wave astronomy when the sources are obscured by dust or molecular clouds.
\end{abstract}


\section{Introduction}
At the present time we witness an ongoing dedicated search for gravitational waves from various astrophysical sources using the ground based laser interferometers \cite{ligo}. Since the sensitivity of these instruments is at the upper bound of assumed intensities coming from realistic sources it might be interesting to estimate the damping effects of various dust formations on the propagating gravitational waves.

In the past several papers (for review see \cite{grishchuk}) dealt with the effects of media on gravitational wave but their main concern was the modification of dispersion relation and not the possible influence on the amplitude. Two different models for the medium most often used were: medium composed of deformable ``molecules'' with internal structure \cite{grishchuk,szekeres} giving rise to anisotropic pressures or free particles with rare collisions described by kinetic theory \cite{lindquist}. To describe the damping in the second model the rate of particle collisions has to be addressed giving rise to the imaginary part of the refractive index \cite{polnarev}. The damping was also studied in viscosity approximation \cite{esposito}. In these treatments the damping was negligible due to the slow accumulation of the phase shift that is important for the damping effect only at the values close to $2\pi$.

In sections 2 and 3 the damping is derived using a straightforward computation without the assumption regarding particle collisions. The interaction of gravitational waves and matter (represented by dust cloud) is derived based on the following picture. The incoming gravitational wave produces periodic oscillation within the ``molecules'' in the dust cloud computed using geodesic deviation equation (hence we use the first of the above mentioned models for the medium). In the second step these oscillations themselves produce gravitational waves derived using multipole approximation. Finally we compose them with the original waves propagating through the cloud from a distant source. We use the term dust because we assume no interaction between the ``molecules''. It means that the model cannot describe effects connected with collective phenomena, on the other hand it can still accommodate selfgravitation of the dust cloud if it can be encoded in the background metric.

In section 2 we assume that the wave is approximately planar at the region of the cloud and the background might be described using Minkowski metric so the source of gravitational wave has to be far enough.
In section 3 we generalize the procedure to curved background and treat the example of a spherical shell of dust around central source of gravitational waves. To arrive at the solution in this case the generalization of quadrupole momentum tensor to this specific problem is given.
Section 4 briefly summarizes results of a previous work \cite{ehlers,ehlers2} and mentions its seemingly unfavorable implication. However it is shown that the result concerns orders of magnitude that are not decisive for any concrete implications.

\section{Derivation for planar waves in Minkowski background}
We start with the simplest case where the geometry in the neighbourhood of the cloud is assumed to be flat. First, we recall the behavior of particle separations in the flat background (that we use for simplification in the region of the cloud) under the influence of gravitational wave (determined by geodesic deviation). However we will generalize the formula to the situation where the reference center of the ``molecule'' is geodesic while the particle (whose separation from the center we want to determine) is additionally sitting in a harmonic potential (described by ``spring constant'' $k>0$) centered at the initial position and feeling dissipative effect of its environment (described by ``damping constant'' $b>0$). At the same time, this is the most straightforward generalization of the second order differential equation using only linear terms. Generalization of geodesic deviation equation to include nongravitational forces is usually called relative deviation. We suppose that the particle is in rest before the wave arrives and therefore only use the so called ``steady state'' solution of driven, damped oscillator equation that serves for generalization of the geodesic deviation equation
\begin{equation}\label{gw-deviation}
x_{B}^{k}(\tau)=x_{B}^{j}(0)[\delta_{j}^{k}+{\textstyle\frac{1}{2}}Ch^{\TT}{}^{k}_{j}(\tau)]|_{A}\ \quad\quad C=\frac{\omega^2-k+ib\omega}{{\textstyle \frac{(\omega^2-k)^2}{\omega^2}}+b^2}\ ,
\end{equation}
where $\omega$ is the frequency of gravitational wave and $i$ is a complex unit. The equation (\ref{gw-deviation}) expresses the position of particle B in proper reference frame of particle A lying at the origin. Transverse-traceless (TT) calibration was used to compute the Riemann tensor components needed in geodesic deviation equation. It should be noted that there exists TT system which agrees to first order in perturbation with the proper reference frame. In the following we will use exactly this system for further computations.

Next, we need to know how these periodic movements generate gravitational waves. For this purpose we use the reduced quadrupole moment
\begin{equation}
\Ibar_{jk}=\int \rho (x_{j}x_{k}-\textstyle{\frac{1}{3}}r^2\delta_{jk})\d^{3}x\ .\end{equation}
We compute this expression using the equation (\ref{gw-deviation}). Dropping the particle labels and assuming that the density $\rho$ of the dust stays approximately constant throughout the passage of the wave (cross-section changes only in the second order of perturbation) the result computed to linear terms in perturbation is the following
\begin{equation}\label{quadrupole}
\Ibar_{jk}=\rho\int_{\bar{V}}\left[(\bar{x}_{j}\bar{x}_{k}-\textstyle{\frac{1}{3}}\bar{r}^2\delta_{jk})+{\textstyle\frac{1}{2}}C(\bar{x}_{j}\bar{x}_{m}h^{\TT}{}_{k}^{m}+\bar{x}_{k}\bar{x}_{m}h^{\TT}{}_{j}^{m}-\textstyle{\frac{2}{3}}\bar{x}_{m}\bar{x}^{l}h^{\TT}{}_{l}^{m}\delta_{jk})\right]\d^{3}\bar{x}\ ,
\end{equation}
where $\bar{x}$ denotes the positions before the wave arrival. It indicates, that generally, we cannot neglect the influence of matter on passing gravitational radiation, since the generated intensity depends on second time derivative of reduced quadrupole moment. We can see that this intensity would be produced by the terms in the second round bracket in equation (\ref{quadrupole}). 

The generated perturbations ${}_{g}h$ in TT frame have the following form \cite{MTW}
\begin{equation}\label{TT-waves}
{}_{g}h_{jk}^{\TT}(t-r)=\frac{2}{r}\ddot{\Ibar}_{jk}^{\TT}(t-r)+O(1/r^2)\ ,
\end{equation}
where 
\begin{equation}
\Ibar_{jk}^{\TT}=P_{j}^{a}\Ibar_{ab}P_{k}^{b}-\textstyle{\frac{1}{2}}P_{jk}P^{ab}\Ibar_{ab}
\end{equation}
and $P_{ab}=\delta_{ab}-n_{a}n_{b}$ is the projection operator to the subspace transverse to unit radial vector $n_{a}$.

\subsection{Example with cuboid}
Let us show the computation for one very simple situation. Suppose the cubical dust cloud has approximate sizes $X$,$Y$ and $Z$ (with respect to Cartesian coordinates) in the reference frame with origin in the clouds center and the $z$-axis pointing to the distant source of gravitational waves on one side and to distant observer on the other side. We assume that the wave from external source is approximately planar in the region of the cloud. We compute the second time derivative of highest order terms for the relevant components of reduced quadrupole moment in TT gauge (with $n^{a}=(\partial_{z})^{a}$) for a ``molecule''(region where the driven, damped oscillator model described above might be applied) of linear dimension $a$ obtaining
\begin{eqnarray}
\ddot{\Ibar}_{xx}^{\TT}&=&\textstyle{\frac{1}{6}}C\rho a^{5}\ddot{h}^{\TT}_{xx}\nonumber\\
\ddot{\Ibar}_{yy}^{\TT}&=&\textstyle{\frac{1}{6}}C\rho a^{5}\ddot{h}^{\TT}_{yy}\\
\ddot{\Ibar}_{xy}^{\TT}&=&0\nonumber
\end{eqnarray}
where $h^{\TT}(z=0)\sim e^{-i\omega t}$ is a planar perturbation in TT gauge from distant source (we have chosen the $\oplus$ polarization mode for simplicity). By superposing the generated gravitational waves (\ref{TT-waves}) from all molecules in a given layer of dust cloud in the $xy$-plane we obtain a planar wave (neglecting the effects of finite dimensions of the cloud)
\begin{equation}\label{gen-wave}
{}_{g}h^{\TT}=\frac{4\pi}{i\omega}\sigma\ddot{\Ibar}^{\TT}=-\frac{2\pi C}{3i}\omega\sigma\rho a^{5}{h}^{\TT}
\end{equation}
where the planar density of ``molecules'' $\sigma$ can be computed as ${a^{-2}}$ for a dense medium and equality $\ddot{h}^{\TT}=-\omega^{2}{h}^{\TT}$ was used. Now we add the generated wave (\ref{gen-wave}) to the original perturbation ${h}^{\TT}$. This process will happen in subsequent layers with efficiency described by parameter $\kappa=<0..1>$ (roughly speaking it would happen in every $1/\kappa$ slice) which gives the following exponential decay of the amplitude of gravitational perturbation as it travels through the cloud
\begin{equation}\label{cuboid-damping}
h_{A}=h_{B}\,\e^{-\frac{2\pi\kappa C}{3i}\omega\sigma\lambda Z\rho a^{5}}\ ,
\end{equation}
where $\lambda$ is linear density of molecules in the direction of propagation and we have omitted coordinate indices, $h_{B}$ is the amplitude before the cloud and $h_{A}$ after it. We may define a volume number density of ``molecules'' as $\tilde{\rho}=\sigma\lambda$.

Since the exponential factor in (\ref{cuboid-damping}) is complex the overall effect is twofold. The imaginary part is responsible for the phase shift and the real part for the possible damping. The real part of the factor can be written as $-\frac{2\pi\kappa \Im (C)}{3}\omega\tilde{\rho} ZM a^{2}$, where $M$ is the mass of the ``molecule''. This number is non-positive and therefore really describes exponential decay of amplitude.

  Although this picture seems to be pretty naive there is direct correspondence between above mentioned second time derivatives of quadrupole moment and perturbations of stress energy tensor representing anisotropic stresses that generate gravitational waves via standard linearized Einstein equations (for detailed discussion see \cite{grishchuk}).

\section{``Spherical waves'' and the dust shell}
Now we are going to a more complex example where the approach from the previous section cannot be applied straightforwardly. As a background we use the Vaidya--(anti-)de Sitter metric in stereographic coordinates $(u,r,\eta,\xi)$
\begin{equation}
\d s^2=(-1+\frac{2m(u)}{r}+H^2r^2)\d u^2+2\d u\d r+\frac{r^2}{p^2}(\d\eta^2+\d\xi^2)\ ,
\end{equation}
where $p(\eta,\xi)=1+1/4(\eta^2+\xi^2)$ and $H=\sqrt{\Lambda/3}$ (with $\Lambda$ being a cosmological constant). To determine the form of gravitational waves we can use the results of paper \cite{svitek-WKB} where the background is a general Robinson-Trautman spacetime with cosmological constant (describing also the above metric as a special case) and high-frequency approximation developed by Isaacson \cite{isaac} is used. The wave vector is $k_{\mu}=({\phi}_{,u},0,0,0)$ and we specialize to the $\oplus$ polarization mode. So the only nonzero elements of the perturbation tensor are the following
\begin{equation}
 h_{\eta\eta}=A\frac{r}{\sqrt{2}p^2}\exp({i\phi(u)})=-h_{\xi\xi}\ ,
\end{equation}
with $A$ being an amplitude and $\phi$ the phase. The effect will be analyzed with respect to the stationary observer near infinity with four-velocity $\mathbf{v}=(\dot{u},0,0,0)$. For zero cosmological constant this observer is almost geodesic. Once again we will employ the geodesic deviation equation (in a more general form \cite{MTW})
\begin{equation}\label{geodev}
\nabla_{\mathbf{v}}\nabla_{\mathbf{v}}n^\alpha=-R^{\alpha}_{\beta\gamma\delta}v^{\beta}n^{\gamma}v^{\delta}\ .
\end{equation} 
The deviation vector $n^\alpha$ connecting neighboring geodesics has components in the $\partial_{\eta}$ and $\partial_{\xi}$ directions only. Assuming that the spherical dust shell that interacts with the gravitational wave starts the free fall from rest ($\frac{\partial n^{\alpha}}{\partial u}=0$) the right-hand side of (\ref{geodev}) could be written as $\frac{\d^2 n^{\alpha}}{\d u^2}\dot{u}^2$ and $\dot{u}=(-g_{uu})^{-1/2}$ from four-velocity normalization. The Riemann tensor could be decomposed into background part $R^{(0)}$ and contribution induced by perturbations $R^{(1)}$ which is the dominant one in the high-frequency approximation \cite{isaac}
\begin{equation}
 R^{(1)}_{\alpha\beta\gamma\delta}=2h_{[\gamma|[\beta;\alpha]|\delta]}\quad\Longrightarrow\quad R^{(1)}_{u\eta u\eta}=-2^{-3/2}\frac{rA}{p^2}\frac{\d^2}{\d u^2}\exp({i\phi(u)})=-R^{(1)}_{u\xi u\xi}\ .
\end{equation}
Now we can integrate the simplified equation (\ref{geodev}) (assuming that $R^{(0)}$ changes very slowly) with added effect of non-gravitational forces (i.e., when the particle position satisfies $\nabla_{\mathbf{v}}\nabla_{\mathbf{v}}\mathbf{n}=-R(\mathbf{v},\mathbf{n})\mathbf{v}-b\nabla_{\mathbf{v}}\mathbf{n}-k(\mathbf{n}-\mathbf{n}_{0})$, generalizing the relative deviation used in the flat case) to obtain
\begin{equation}\label{spher-dev}
 n^{\eta}=\left(1+\frac{R^{(0)}{}^{\eta}_{u\eta u}}{k g_{uu}}+2^{-3/2}\frac{rA\tilde{C}}{p^2}g^{\eta\eta}\exp({i\phi(u)})\right)n_{0}^{\eta}\ ,
\end{equation}
where $\tilde{C}=\frac{(\phi_{,u})^2+g_{uu}k+i\sqrt{-g_{uu}}b\phi_{,u}}{\frac{((\phi_{,u})^2+g_{uu}k)^2}{(\phi_{,u})^2}-g_{uu}b^2}$.

Next we would need a fully relativistic generalization of the quadrupole moment defined using integrals over the source. However this seems to be a very tricky question and the only agreed definition relies on identifying the corresponding contribution in the multipole expansion of the field generated by the source which is not useful in our case. For example in the case of rigidly rotating axially symmetric star the multipole moments were numerically computed using iteration techniques and Green function method by Ryan \cite{ryan}. But the method is too complicated for our purpose so we will try to generalize the flat-space formula since our two-space where the deviation vector resides is conformally flat (to remedy the problem with coordinates without simple length interpretation we use tensor with both types of indices). Our prescription for a small element on the pole is the following ($A, B$ take values $\eta$ and $\xi$)
\begin{equation}
 \Ibar_{A}^{B}=\rho\underset{K(r_{0},r_{0}+\triangle r)\cap \triangle\Omega}\int[n_{A}n^{B}-1/3r^2\delta_{A}^{B}]g_{rr}\frac{r^2}{p^2}\d r\d \eta\d \xi\ ,
\end{equation}
where $\triangle\Omega=(\{-\triangle \eta/2,\triangle \eta/2\}\times\{-\triangle \xi/2,\triangle \xi/2\})$, $\triangle \eta=\triangle \xi=a/r_0$ and the metric coefficient $g_{rr}$ comes from Schwarzschild-like form of the metric. Integration is over the intersection of a shell between the indicated radii and small spatial angle. Using the integral mean value theorem, setting $n_{A}=g_{AB}n^{B}$, expressing the deviation vector as $n^{\eta}_{0}=\eta$ ($n^{\xi}_{0}=\xi$) and using (\ref{spher-dev}) we obtain in the highest order (already using the \TT-gauge)
\begin{equation}
\frac{\d^2}{\d u^2}\Ibar^{\TT}{}_{\eta}^{\eta}\doteq 2^{-3/2}\left(1+\frac{R^{(0)}{}^{\eta}_{u\eta u}}{k g_{uu}}\right)\rho A\tilde{C}g_{rr}\triangle r\,r_{0}^{3} \frac{\d^2}{\d u^2}(\exp({i\phi(u)}))\int_{\triangle\Omega}\frac{\eta^2+\xi^2}{p^{4}}\d \eta\d \xi\doteq
\end{equation}
\begin{equation}\nonumber
{}\doteq-\frac{2^{-5/2}}{3}\left(1+\frac{R^{(0)}{}^{\eta}_{u\eta u}}{k g_{uu}}\right)\rho A\tilde{C}g_{rr}\triangle r\,r_{0}^{-1}a^4(\phi_{,u})^{2}\exp({i\phi(u)})\ +O(a^5),
\end{equation}
where we have neglected the terms containing second derivative of phase in accordance with \cite{isaac} and $g_{rr}$ might be approximately set to one when we are not close to horizons. In the same way one obtains $\frac{\d^2}{\d u^2}\Ibar^{\TT}{}_{\xi}^{\xi}=-\frac{\d^2}{\d u^2}\Ibar^{\TT}{}_{\eta}^{\eta}$ and $\frac{\d^2}{\d u^2}\Ibar^{\TT}{}_{\eta}^{\xi}=0$. Now, we need to use equation (\ref{TT-waves}) to obtain gravitational waves generated by the element and subsequently add the contributions from a given layer. In the spherical case we count (in integration) only those elements that are ``visible'' on the surface of the spherical layer from the point on the next layer (for the north pole it means $\theta=<0..\arccos(\frac{r_{0}}{r_{0}+\triangle r})>$), therefore not taking into account curved geometry where the waves propagate. When neglecting the term periodically dependent on the square root of distance (which corresponds to discarding the term related to boundary in the flat case) we obtain
\begin{equation}
{}_{g}h^{\TT}{}_{\eta}^{\eta}=\frac{2\pi i r_{0}}{\phi_{,u}r}\sigma\frac{\d^2}{\d u^2}\Ibar^{\TT}{}_{\eta}^{\eta}\ ,
\end{equation}
where $\sigma$ is the surface density of the ``molecules''. For the covariant metric perturbation one may then write
\begin{equation}
h^{\TT}_{\eta\eta}\left(r_{0}+(j+1)\triangle r\right)=\left[1+g_{rr}\vert_{(r_{0}+j\triangle r)}K\triangle r\right]h^{\TT}_{\eta\eta}(r_{0}+j\triangle r)\ ,
\end{equation}
where
\begin{equation}
K=-\frac{8\pi i}{3}\left(1+\frac{R^{(0)}{}^{\eta}_{u\eta u}}{k g_{uu}}\right)\sigma\rho \tilde{C}a^4\phi_{,u}\ .
\end{equation}
Setting $\triangle r (=\lambda^{-1})=\frac{r_1-r_0}{N}$ (with $N>>1$ being the number of layers and $r_0,r_1$ the inner and outer radius of a dust shell, respectively) and assuming that the wave decreases with efficiency $\kappa$(as in the planar model) we get for the intensities under inner radius $h_{B}$ and above outer radius $h_{A}$
\begin{equation}
 h_{A}=h_{B} \prod _{j=0}^{N}\left[1+\frac{K{\triangle r}}{\left( 1-2\,{\frac {m}{{r_0}+j{\triangle r}}}-{H}^{2} \left( {r_0}+j{\triangle r} \right) ^{2} \right)}\right]\doteq h_{B} \left[1+{\frac { \left( N+1 \right) {{r_0}}K{\triangle r}}{{r_0}-2\,m-{H}^{2}{{r_0}}^{3}}}+\right.
\end{equation}
\begin{equation}\nonumber
\left.+\frac{1}{2}\,{\frac {K \left( N+1 \right) N \left( r_{0}^2 K-2\,m+2\,{H}^{2}{{r_0}}^{3}\right) {{\triangle r}}^{2}}{ \left({r_0}-2\,m-{H}^{2}{{r_0}}^{3}\right) ^{2}}}+O \left( {{\triangle r}}^{3} \right) \right.
\end{equation}
We can check that this result has at least the same length dimensionality and similar dependence on frequency as for our cuboid case on Minkowski which might give some vindication to the calculation used in the curved space.

\section{Perturbation viewpoint}
In 1987 Ehlers, Prasanna and Breuer \cite{ehlers} investigated the propagation of small-amplitude gravitational waves through
pressureless matter ('dust') using the linearization of Einstein equations for dust
\begin{equation}
g_{\mu\nu}, \rho, u_{\mu} \longrightarrow g_{\mu\nu}+\varepsilon \delta g_{\mu\nu}, \rho + \varepsilon \delta \rho, u_{\mu}+\varepsilon \delta u_{\mu}\ .
\end{equation}
They have used the WKB method to study the locally plane, linearized perturbations of an arbitrary background dust spacetime asymptotically for small wavelengths (thus assuming the high frequency character of the perturbations)
\begin{equation}
\delta g_{\mu\nu}=\Re[\e^{\frac{i}{\varepsilon}S(x)}f_{\mu\nu}(x,\varepsilon)]\ .
\end{equation}
Finally, they have cast the resulting system of equations into the form where $\delta g_{\mu\nu}$ serves as a source for generating perturbations $\delta \rho$ and $\delta u_{\mu}$.

However, they have arrived at the conclusion that in the leading order no perturbations of density or vorticity are generated
\begin{equation}\label{density-perturbation}
\delta \rho=0 \quad \delta u_{\mu}=0\ .
\end{equation}
This makes it impossible to alter the passing gravitational wave in the second step (the medium in this model is unchanged in the first place). This result would not allow the damping of gravitational wave seen in the previous sections. But one should also keep in mind that the model from sections 2 and 3 relies on an internal structure of the ``molecules'' (although their collection behaves globally like dust) and therefore is not directly comparable to the perturbation picture described above.
 
Anyway, careful reading of the paper \cite{ehlers} (and of a more recent paper \cite{ehlers2} dealing with the perfect fluid case) reveals that in fact the order in which the right-hand sides of equations (\ref{density-perturbation}) are supposed to vanish is $\varepsilon^{-2}$. But this is necessary condition for the perturbation method itself to be consistent and does not restrict existence of nonzero matter perturbations at the order $\varepsilon$.  

\section{Conclusion}
The above described simple model (presented in sections 2 and 3) might produce measurable effect on the intensity of detectable gravitational waves. The model assumes certain (simple) interaction of particles inside the ``molecules'' forming the cloud so it most closely resembles molecular cloud. The results described in section 4 were shown not to cause any important restriction on the model from the point of view of the perturbative approach.

To give some notion of the strength of this effect we can compute the damping for some astrophysically relevant data. We use the flat model (section 2) with $k=\omega^2$ (maximal damping effect, see (\ref{gw-deviation})), $b=\omega/10$ (to stay safely at underdamped oscillations when the driving force is neglected), $a=10^{-10} m$ (typical small molecule dimension) and efficiency $\kappa=10^{-1}$ (to account for limited internal structure). We will compute the length over which the amplitude of gravitational waves is damped by $10 \%$. For dense medium ($\tilde{\rho} \sim a^{-3}$) this length is $10^{-12} pc$ for frequency of $100 Hz$ (and $10^{-10}$ for $1 Hz$). For average interstellar medium densities of our galaxy ($10^7 m^{-3}$) the lenght is however $10^{10} pc$ for $100 Hz$ ($10^{12} pc$ at $1 Hz$).

The model is to simple and rough at this stage to make some final judgement considering the observational consequences. Some parameters are not set by clear physically founded estimates. The future work will concentrate on providing a more complex model capturing more precisely the structure of the medium. If one is concerned with molecular structures the approach should probably utilize quantum mechanics rather that classical one at the ``microscopic'' level. Also the treatment of the curved background should be more consistent (e.g. the model neglects the effects of curved background in the computation of perturbation produced by given layer).

\section*{Acknowledgement}
I would like to thank Petr Hadrava for initiating my interest in this topic and for useful discussions. This work was supported by grant GACR 202/07/P284 and the Center of Theoretical Astrophysics LC06014. 



\begin{thebibliography}{40}
\bibitem{ligo}
{\tt http://www.ligo.caltech.edu/ \\
http://www.virgo.infn.it/ \\
http://www.geo600.uni-hannover.de/ \\
http://tamago.mtk.nao.ac.jp/}

\bibitem{grishchuk}
Grishchuk L~P and Polnarev A~G 1981, Gravitational waves and their interaction with matter and fields {\em in General Relativity and Gravitation vol. II} ed. A Held 393--434

\bibitem{szekeres}
Szekeres P 1971, Linearized gravitation theory in macroscopic media {\em Ann. Phys.} {\bf 64} 599--630

\bibitem{lindquist}
Lindquist E~W 1966, Relativistic transport theory {\em Ann. Phys.} {\bf 37} 487--518

\bibitem{polnarev}
Polnarev A~G 1972, The interaction of weak gravitational waves with a gas {\em Zh. Eksp. Theor. Fiz.} {\bf 62} 1598--1604

\bibitem{esposito}
Esposito F~P 1971, Absorption of gravitational energy by a viscous compressible fluid {\em Astrophys. J.} {\bf 165} 165--170

\bibitem{svitek-WKB}
Podolsk\'{y} J and Sv\'{\i}tek O 2004, Some high-frequency gravitational waves related
to exact radiative spacetimes 
{\em Gen. Rel. Grav.} {\bf 36} 387--401

\bibitem{isaac}
Isaacson R~A 1968, 
Gravitational radiation in the limit of high frequency I., II.
{\em Phys. Rev.} {\bf 166} 1263--1280

\bibitem{MTW}
Misner C~W, Thorne K~S and Wheeler J~A 1973.
{\em Gravitation} (W.~H.~Freeman: San Francisco).

\bibitem{ryan}
Ryan F~D 1997,Spinning boson stars with large self-interaction {\em Phys. Rev.} D {\bf 55}, 6081

\bibitem{ehlers}
Ehlers J, Prasanna A~R and Breuer R~A 1987, Propagation of gravitational waves through pressureless matter {\em Class. Quantum Grav.} {\bf 4} 253--264

\bibitem{ehlers2}
Ehlers J and Prasanna A~R 1996, A WKB formalism for multicomponent fields and its application to gravitational and sound waves in perfect fluids {\em Class. Quantum Grav.} {\bf 13} 2231--2240

\end{thebibliography}
\end{document}